\documentclass[preprint,aps]{revtex4}
\usepackage{graphicx}
\usepackage{dcolumn}
\usepackage{bm}
\begin{document}
\title{ Soliton solutions in the class of  implicit  difference schemes}
\author{V.V. Makhro}
\email{Makhro@mail.ru} \affiliation{Irkutsk State University,
Bratsk Branch, 665708, Bratsk, prospekt Lenina, 34}

\date{August 1, 2004 }

\begin{abstract}
We announce a detailed numerical investigation for some class of difference schemes, which arises from Euler implicit scheme. Such schemes demonstrate unusual behavior and leads to origin of solitons.   Applications to some nonlinear problems are discussed.
\end{abstract}

\maketitle

For the numerical calculations in mathematical physics, one usually applies standard procedure in which initial differential equation is replaced by system of finite-difference equations in accordance with some difference scheme. However, contrary situations when the problem is formulates initially in the terms of some difference scheme also are known. May be the best example is the numerical FPU experiment. In 1955, Fermi, Pasta, and Ulam took numerical scheme for the chain of harmonic oscillators coupled with a quadratic nonlinearity and investigated how the energy in one mode would spread to the rest \cite{1,2}. They found the system cycled periodically, implying it was much more integrable than they had thought. Only in 1963, Martin Kruskal and Norman Zabusky analytically study the Fermi-Pasta-Ulam heat conduction problem in the continuum limit and find that the Korteweg - de Vries equation governs this system \cite{3}.

In our note, we consider similar situation when the problem with very intriguing solutions, which was formulates for the some different scheme, has not continual analogue yet.
Discussed problem comes from the rounding error analysis of the Euler implicit scheme for the linear diffusion equation

\[
u_{i,j + 1} \left( {1 + 2 \cdot c} \right) + c \cdot u_{i - 1,j + 1}  + c \cdot u_{i + 1,j + 1}  = u_{i,j}
.\]

Let`s replace in this scheme $2$ with some arbitrary coefficient $\lambda$
\begin{equation}
 \label{1}
u_{i,j + 1} \left( {1 + \lambda  \cdot c} \right) + c \cdot u_{i - 1,j + 1}  + c \cdot u_{i + 1,j + 1}  = u_{i,j}
 \end{equation}
and examine numerically how solutions will changes with changing of $\lambda$. Let`s take initial conditions as $u_{\frac{n}{2},1}  = 1$ and boundary conditions as $u_{1,j}  = 0$ and $u_{n,j}  = 0$, where $i \in \left[ {1,n} \right]$ and $j \in \left[ {1,m} \right]$.

\begin{figure}[hbt]
\includegraphics[width=\textwidth]{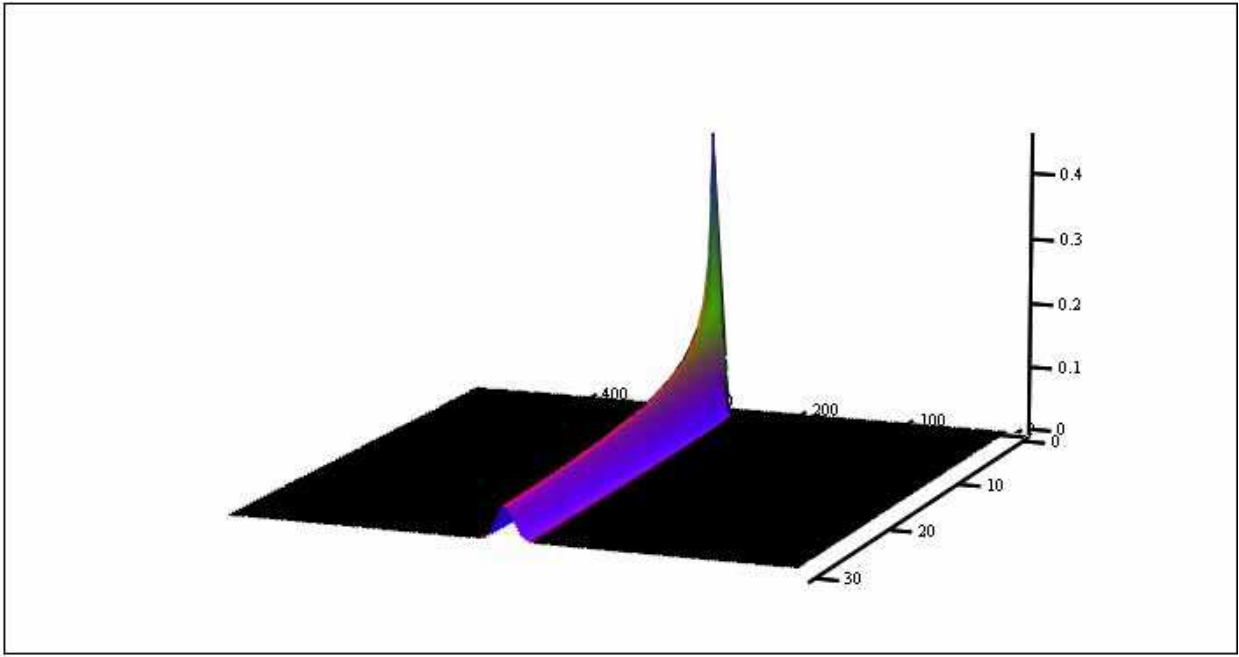}
 \caption{Solution for the linear diffusion equation. $c=-1$, $i \in \left[ {1,500} \right]$, $j \in \left[ {1,30} \right]$}
  \label{Fig1}
\end{figure}

\begin{figure}[hbt]
\includegraphics[width=\textwidth]{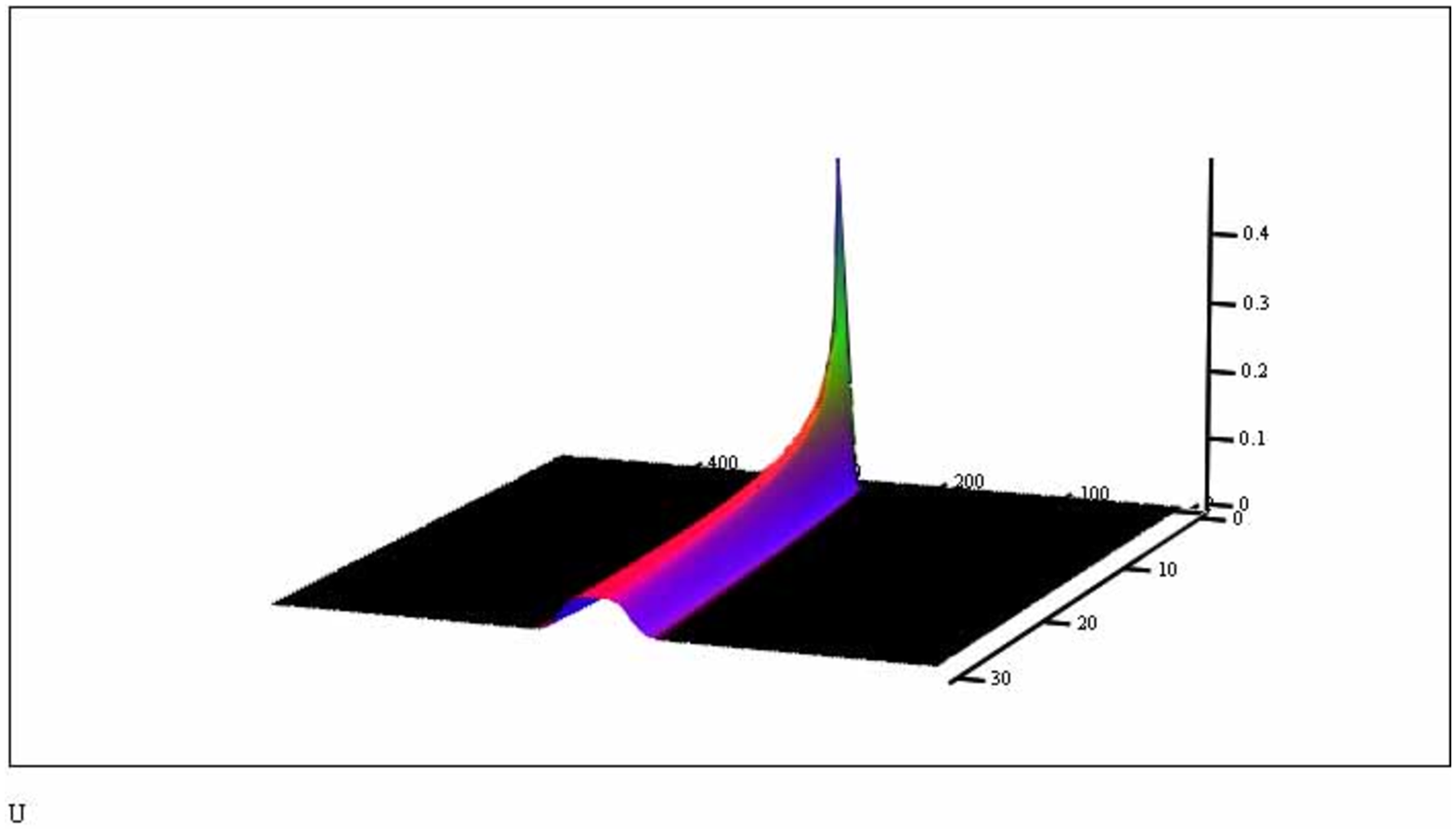}
 \caption{Solution for the linear Schr\"odinger equation, $c=-i$ }
  \label{Fig2}
\end{figure}

Obtained results are as follows.
For $\lambda=2$ and for real $c$ one obtain standard solution for the diffusion equation (Fig. \ref{Fig1}), for the complex $c$ -- usual solution of the linear Schr\"odinger equation (Fig. \ref{Fig2}).

\begin{figure}[hbt]
\includegraphics[width=\textwidth]{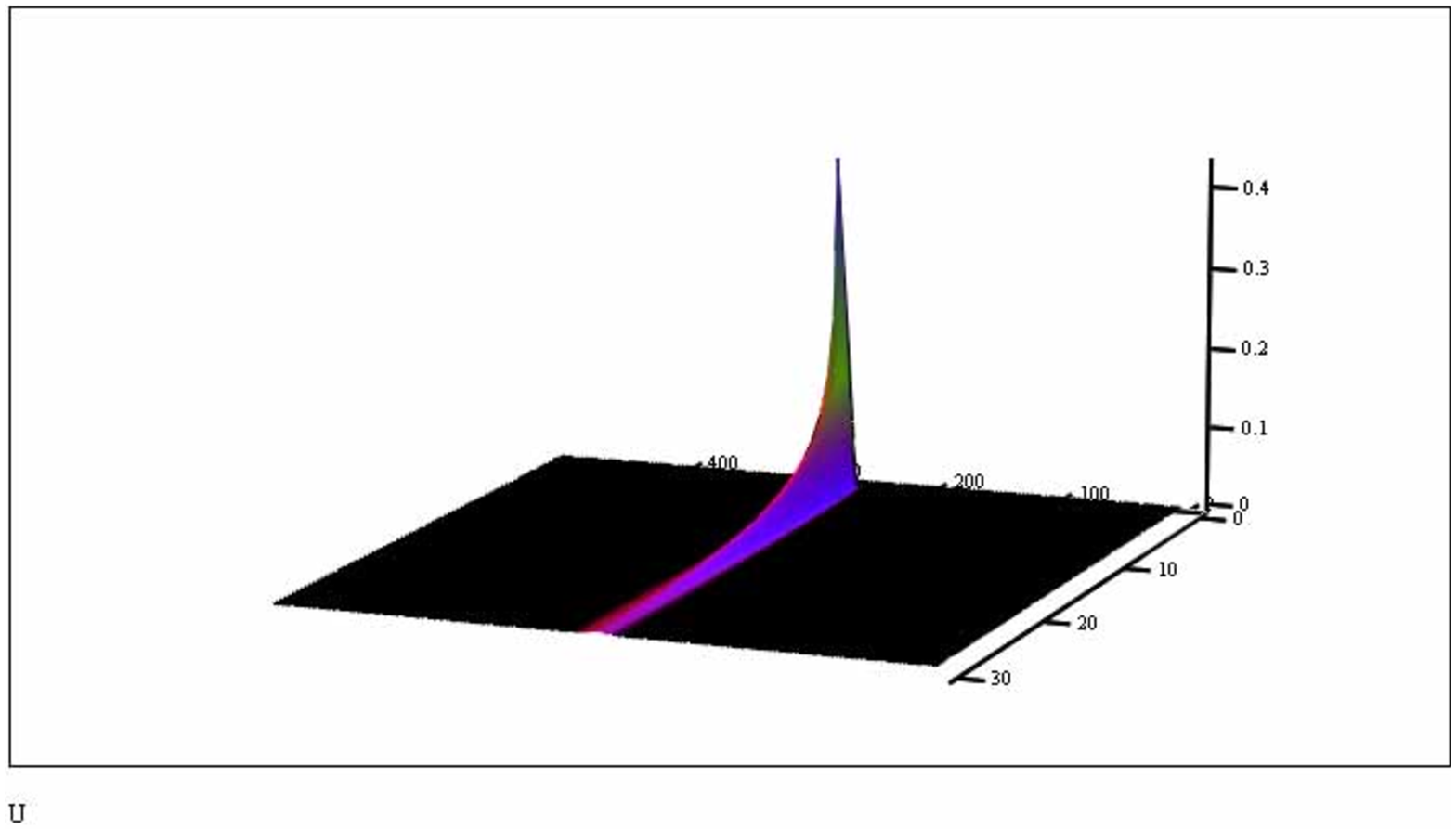}
 \caption{ Solution for the difference scheme with parameters $c=-1$, $\lambda=2.1$}
  \label{Fig3}
\end{figure}

\begin{figure}[hbt]
\includegraphics[width=\textwidth]{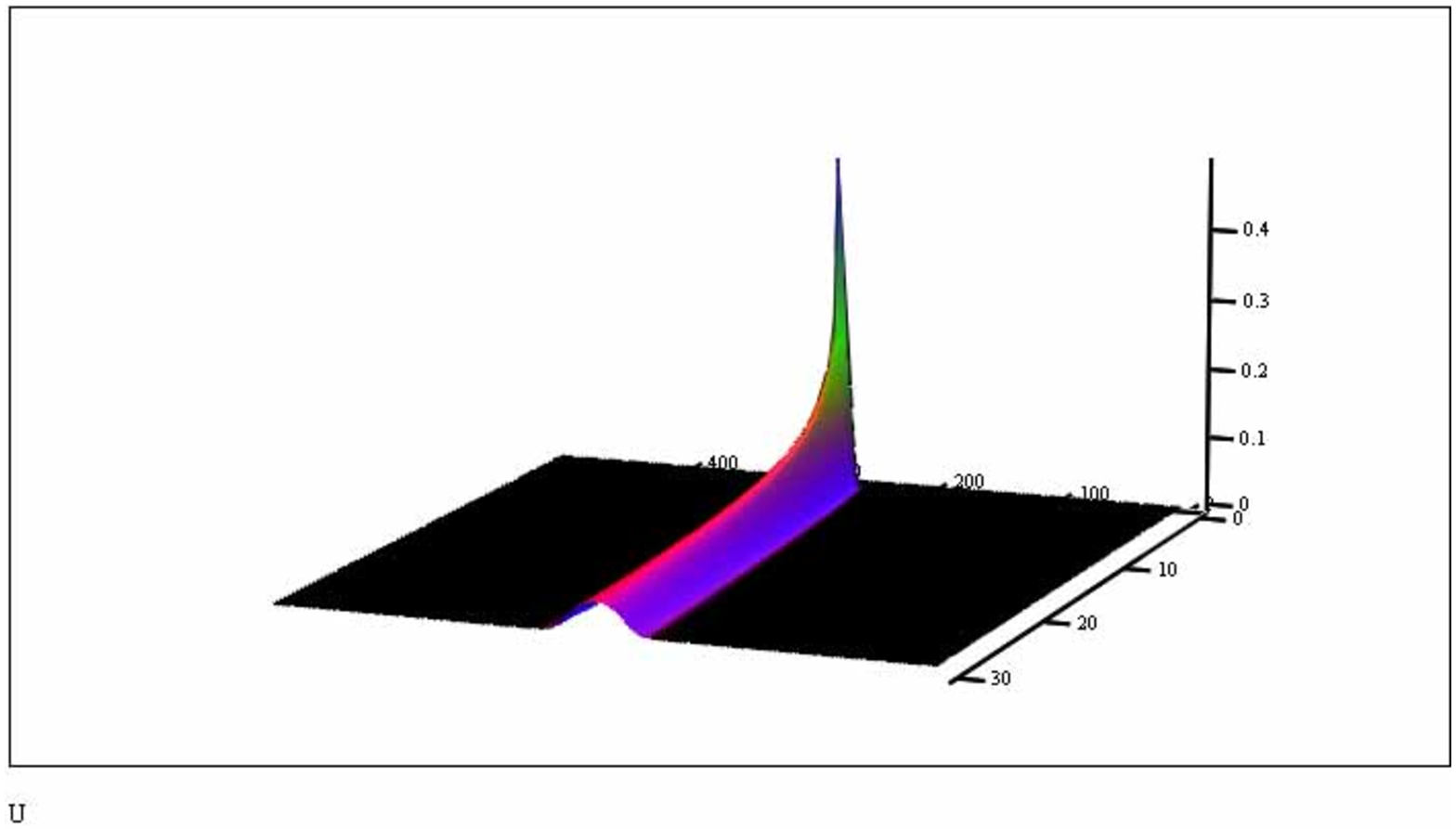}
 \caption{ Solution for the difference scheme with imaginary parameter $c=i$, $\lambda=2.1$}
  \label{Fig4}
\end{figure}

For the $\lambda >2$ solution stay stable for both real and imaginary values of $c$ (Fig. \ref{Fig3} and Fig. \ref{Fig4}).

\begin{figure}[hbt]
\includegraphics[width=\textwidth]{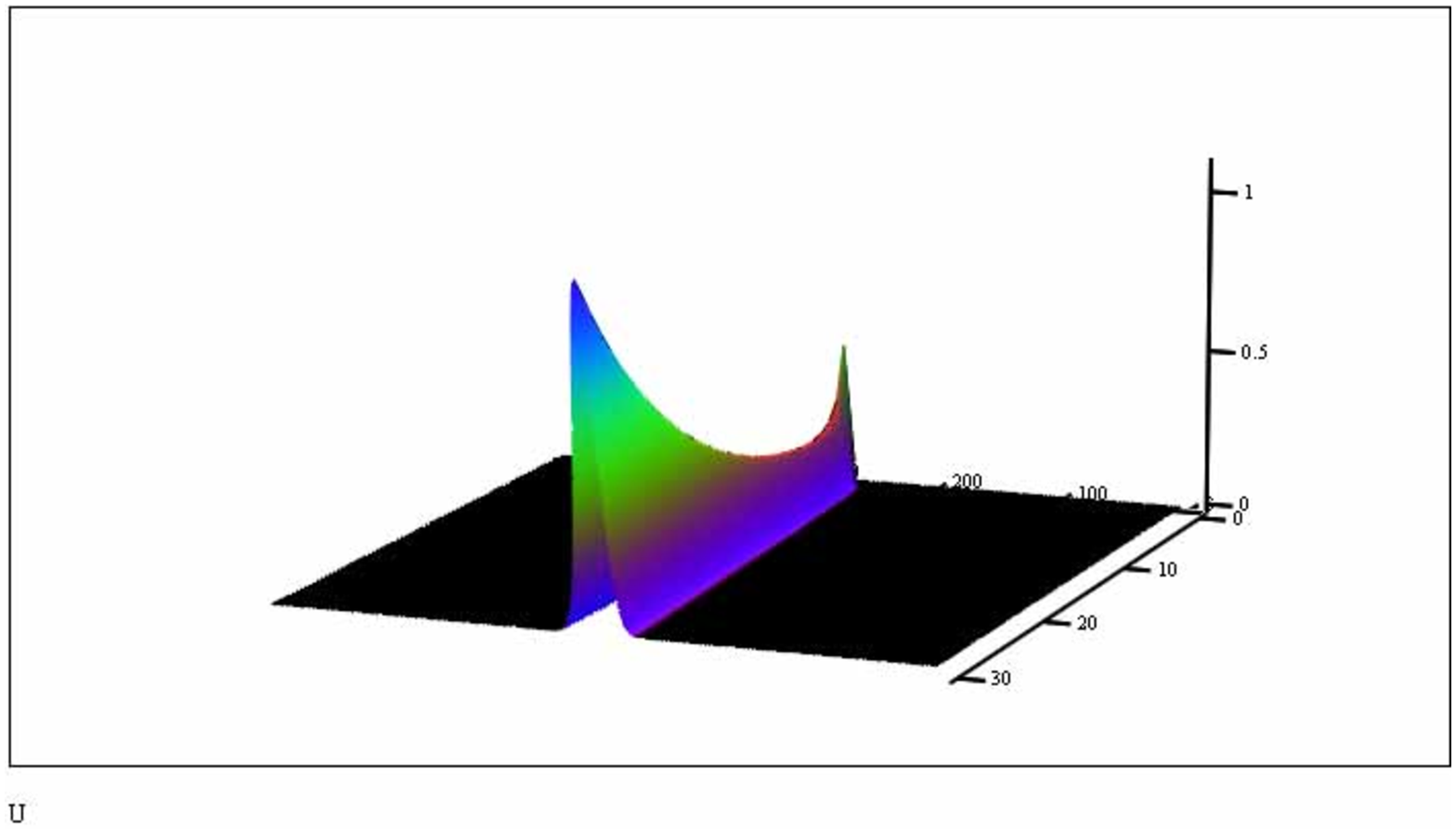}
 \caption{Divergent solution for the difference scheme with $c=-1$, $\lambda=1.9$}
  \label{Fig5}
\end{figure}

However, results for $\lambda<2$ seems to be very different. For all real $c$ the family of schemes (1) gives divergent solutions (\ref{Fig5}). 

\begin{figure}[hbt]
\includegraphics[width=\textwidth]{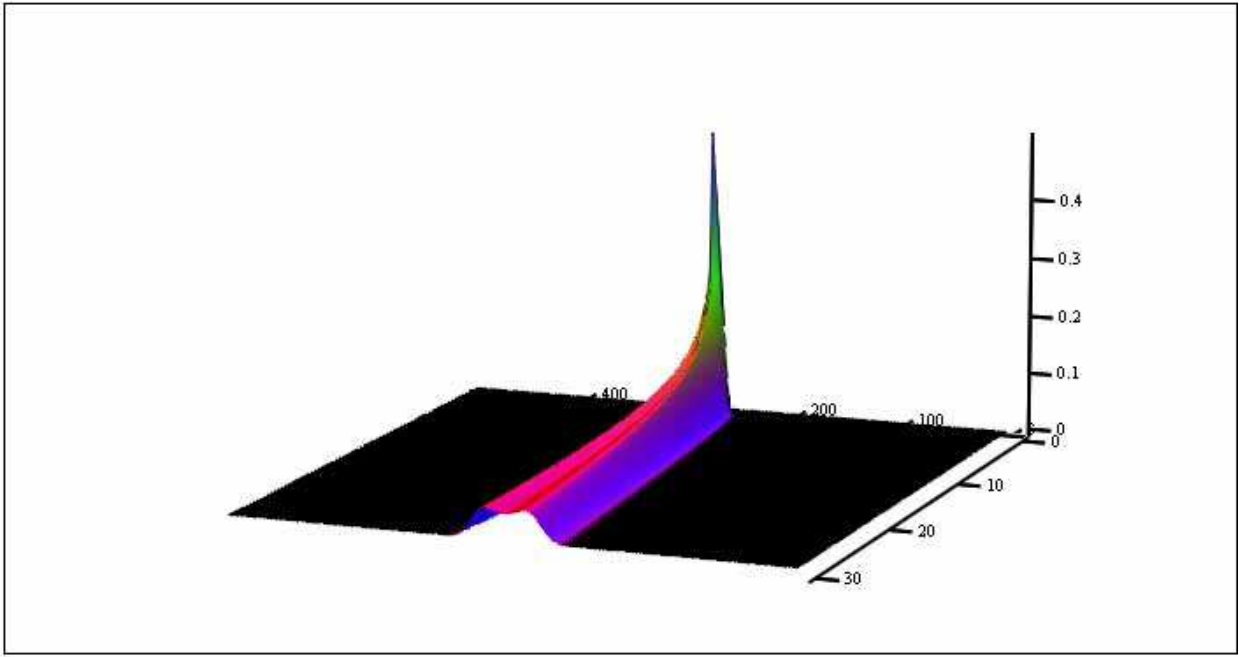}
 \caption{Soliton-like solution for $c=-i$, $\lambda=1.9$}
  \label{Fig6}
\end{figure}

\begin{figure}[hbt]
\includegraphics[width=\textwidth]{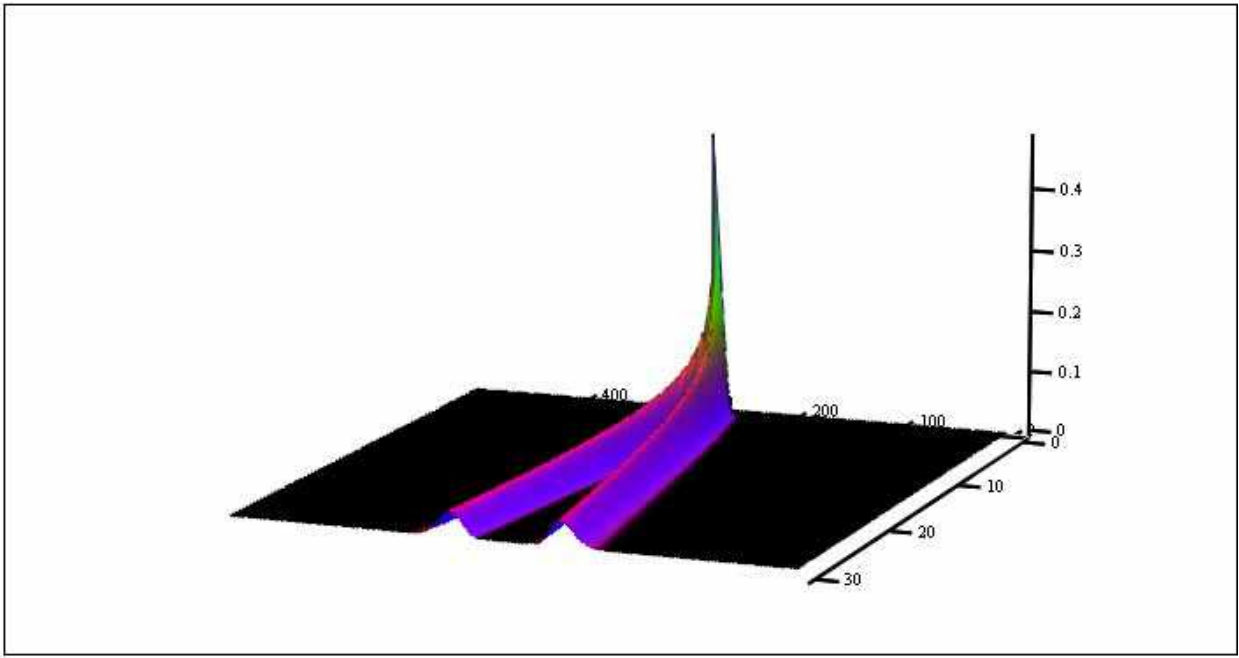}
 \caption{Solitons for $c=-i$, $\lambda=1$}
  \label{Fig7}
\end{figure}

\begin{figure}[hbt]
\includegraphics[width=\textwidth]{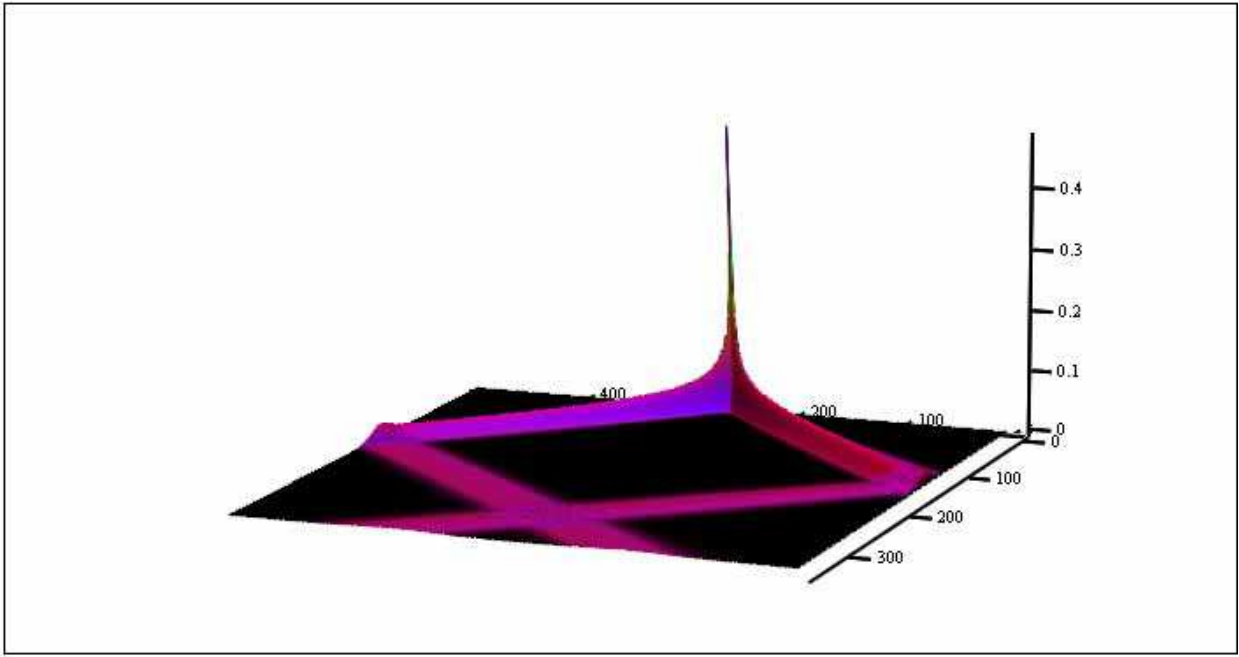}
 \caption{ Scattering of solitons, $c=-i$, $\lambda=1$, $m=350$}
  \label{Fig8}
\end{figure}

More intriguing solutions arises in the case of imaginary $c$. Namely, in this situation for all real values from interval $\lambda  \in \left( {0,2} \right)$ soliton-like formations appear. Some results for this case we present in Fig. \ref{Fig6} -- Fig. \ref{Fig8}. In Fig. \ref{Fig6} where $\lambda=1.9$  for the closest ``time layers`` $j$ solitons has bad resolution.   

\begin{figure}[hbt]
\includegraphics[width=0.75\textwidth]{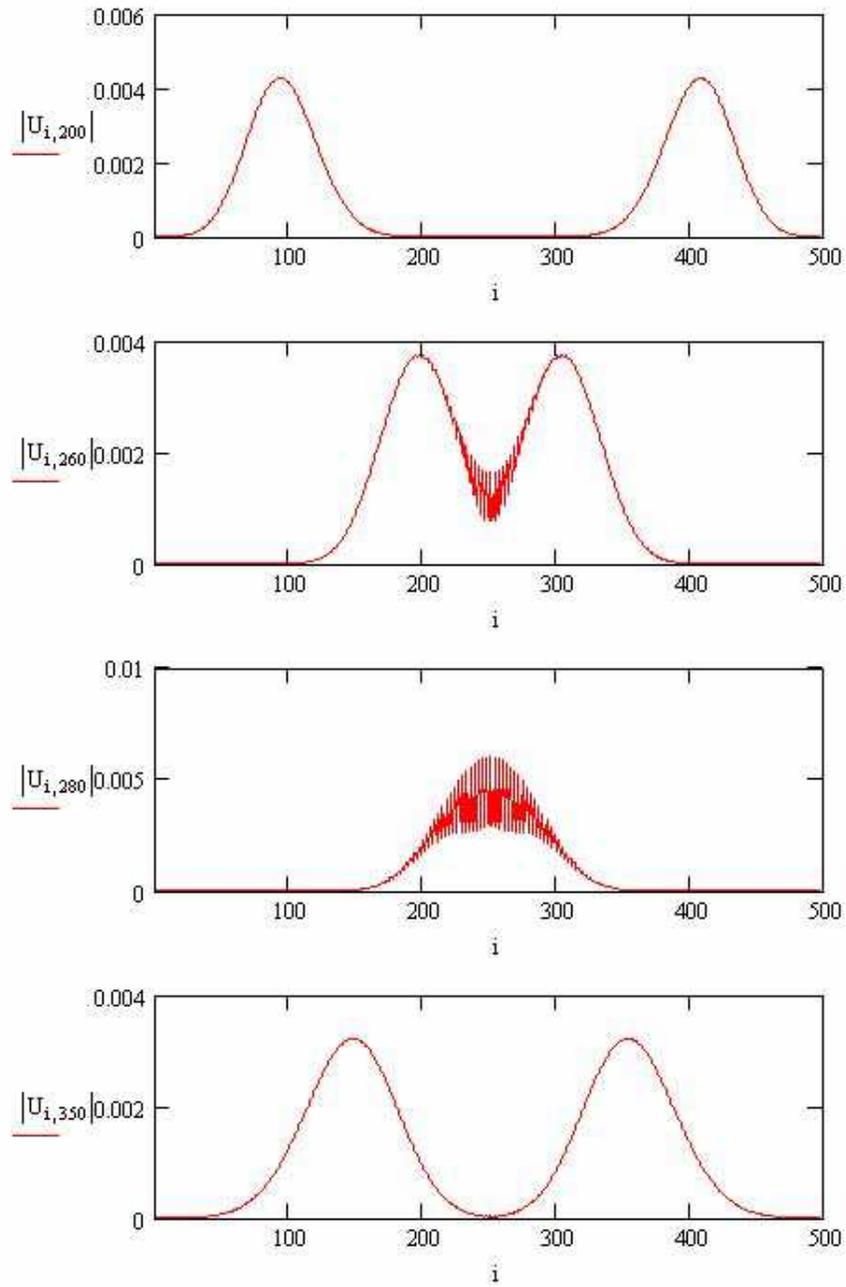}
 \caption{ Scattering in details, see previous figure}
  \label{Fig9}
\end{figure}

\begin{figure}[hbt]
\includegraphics[width=0.75\textwidth]{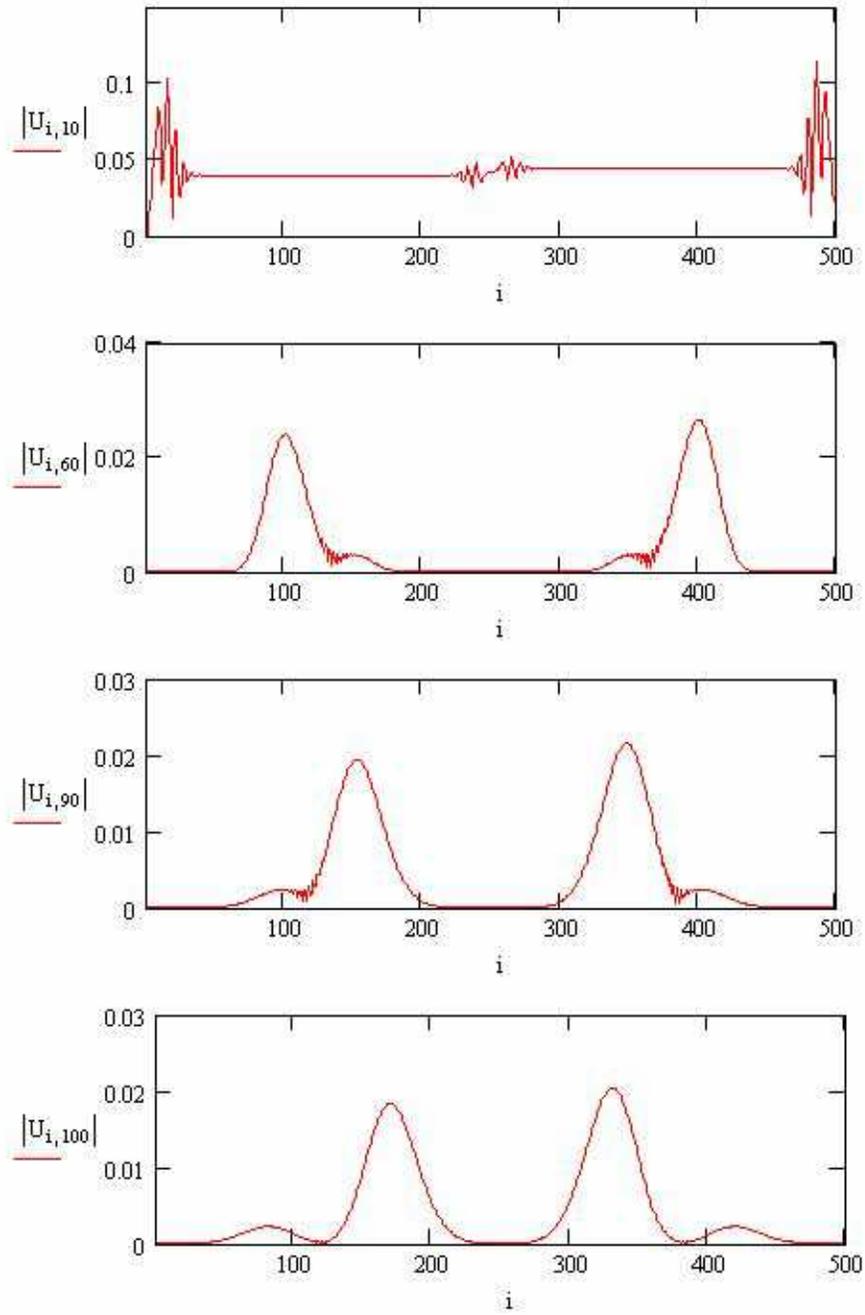}
 \caption{Solitons in difference scheme \ref{1}, $c=-i$, $\lambda=0.5$}
  \label{Fig10}
\end{figure}

\begin{figure}[hbt]
\includegraphics[width=\textwidth]{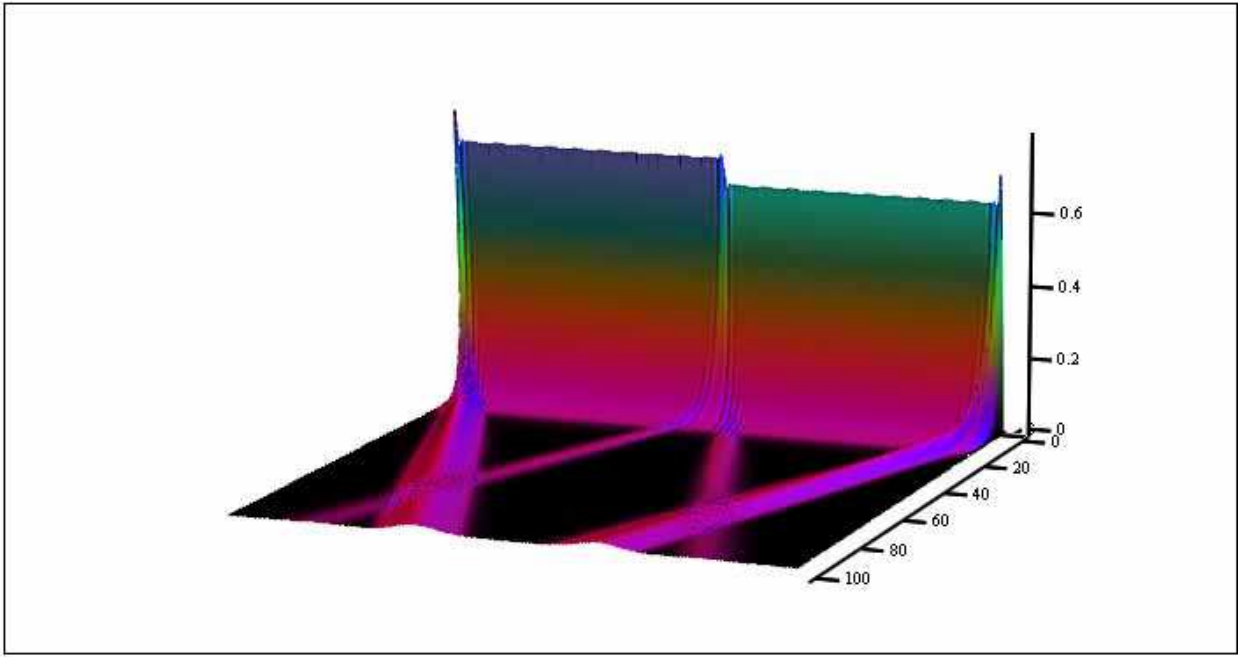}
 \caption{Velocities of solitons, which occurs in the limits of integration range and in its center, are equal}
  \label{Fig11}
\end{figure}

\begin{figure}[hbt]
\includegraphics[width=0.75\textwidth]{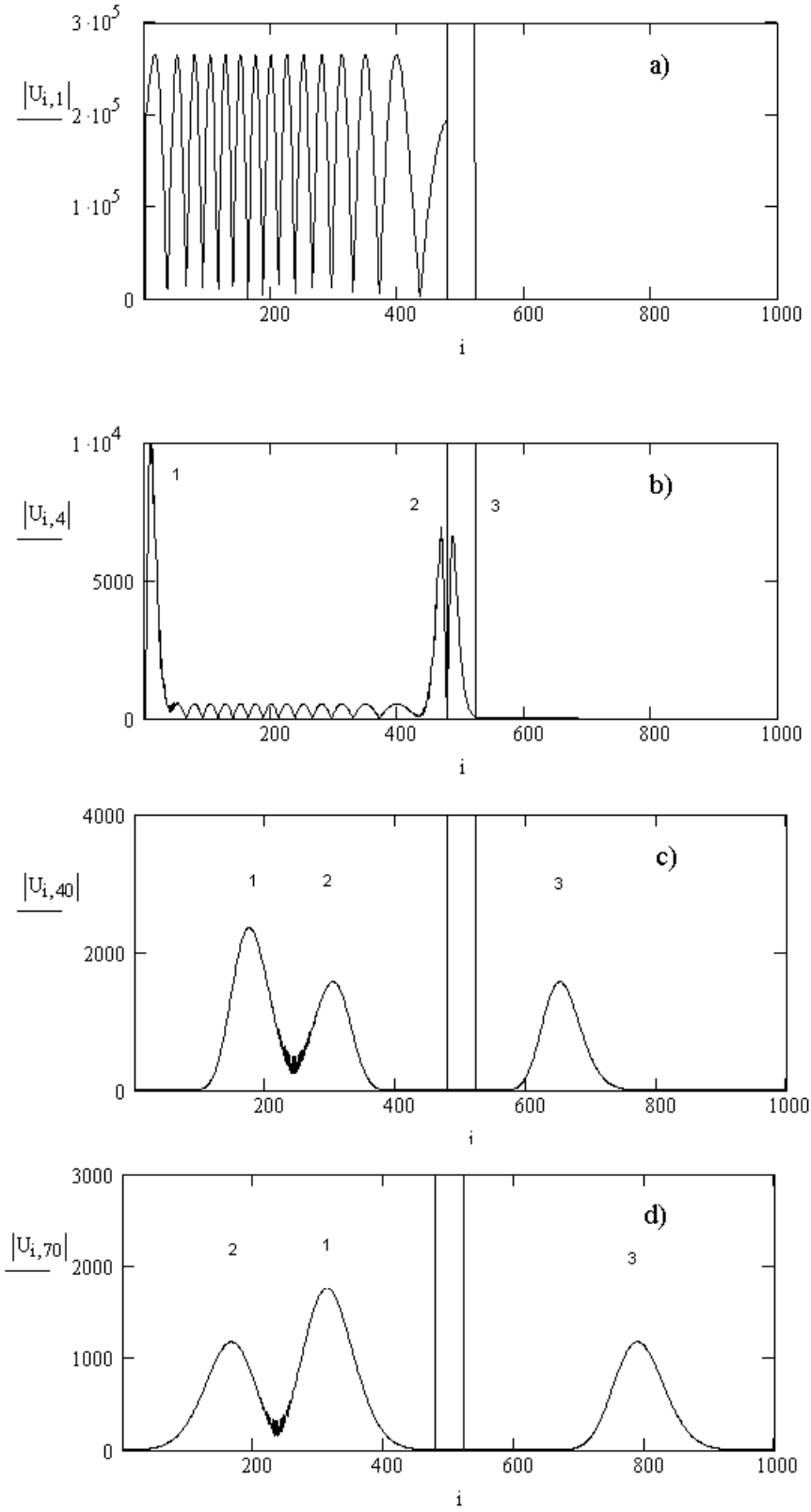}
 \caption{``Soliton-type`` tunneling}
  \label{Fig12}
\end{figure}

\begin{figure}[hbt]
\includegraphics[width=\textwidth]{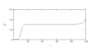}
 \caption{``Tunneling`` rate for $\lambda=1$. Origin of the plateau corresponds with escape of soliton 3 from under the barrier; ending of the plateau corresponds with entering of soliton 3 under the barrier}
  \label{Fig13}
\end{figure}

At the same time solitons velocities and its sharpness grows up with decreasing of $\lambda$. In Fig. \ref{Fig7} we present results of calculations for $\lambda=1$ and the same value of $c$. 

As a whole soliton`s behavior is similar to those of solitons of the Korteweg - de Vries equation. Anyway, its form, character of reflection from the limit points of the integration interval, its scattering – all those factors are very and very similar to KdV solitons (Fig. \ref{Fig8} and Fig.\ref{Fig9}). 
With further decreasing of $\lambda$ soliton`s velocity increased and one can observe its evolution for the deep ``time`` layers $j$.

In the discussed family of difference schemes solitons occurs only in the case when the initial conditions contains the discontinuity of $u$. At that, amplitude is proportional to gradient of $u$ but velocity of solitons stays the same. One can see those features in Fig. \ref{Fig10} and Fig.\ref{Fig11} where initial conditions is $u = 0.9$ for $i \in \left[ {1,\frac{n}
{2}} \right]$ and $u=1$ for $i \in \left[ {\frac{n}
{2} + 1,n} \right]$. 

Presented results of calculations for $\lambda<2$ seems to be amusingly enough but it is a question how those schemes and anomalies correlates with known mathematical (and may be physical) objects. Probably, it is the task of further investigations but some speculations are well-timed now.

Firstly, one can note that soliton`s origin realizes in anyone difference scheme where $\lambda<2$. If $\lambda$ weakly differ from $2$ effects occurs in very deep ``time`` layers (for the big $j$-s) but we believe that effect actual exists. Therefore, such a circumstance one may took in account even for the standard scheme with $\lambda=2$.  

Next, it is incomprehensible now what kind of problems can correspond to schemes with $\lambda<2$ and only rough assumptions are permissible. Though founded solutions similar in appearance with KdV solitons they appear only for imaginary values of $c$. Therefore, one can assume that such solutions relates to Schr\"odinger equation with one or another kind of nonlinearity. It seems to be very stirring.

Indeed, let`s consider the following problem.  Let $\lambda=1$ and initial conditions are given in the form of some harmonic function as it shown in Fig.\ref{Fig12} . Let narrow barrier divides left and right parts of the integration interval. Let`s put initial function inside the barrier and on the right of barrier identical with zero. 

Now, let`s observe evolution of the initial function with the ``time``. In the left border of interval and in the left boundary of barrier solitons appear. Soliton 1 became its motion to the right, soliton 2 – to the left, and soliton 3 – to the right. Solitons 1 and 2 collides with each other and soliton 3 continues its motion. However, it is notable that ratio of integrals of squares of absolute values of considered function stays exactly invariable for the long ``time``. In the case of $\lambda=2$ such a behavior should corresponds to the underbarrier tunneling and above ratio one may consider as tunneling rate (see Fig.\ref{Fig13}). However, in our case it is nothing more than interesting peculiarity.

We hope that further investigations in this direction may be perspective enough.


\begin{thebibliography}{10}
\bibitem{1}
Fermi, E.; Pasta, J.; and Ulam, S. "Studies in Nonlinear Problems, I." Los Alamos report LA 1940, 1955. Reproduced in Nonlinear Wave Motion (Ed. A. C. Newell). Providence, RI: Amer. Math. Soc., 1974.
\bibitem{2}
Tabor, M. "The FUP Experiment." §7.1.b in Chaos and Integrability in Nonlinear Dynamics: An Introduction. New York: Wiley, p. 280, 1989.
\bibitem{3}
Stroboscopic Perturbation Procedure for Treating a Class of Nonlinear Wave Equations, M. D. Kruskal and N. J. Zabusky, 1964. J. Math. Phys. 2, 231-244.
\end{thebibliography}
\end{document}